\documentclass[apj]{emulateapj}
\usepackage{graphicx,natbib,color,amsmath,subfigure}
\usepackage{color}

\shorttitle{Thermal to nonthermal energy partition}
\shortauthors{Altyntsev et al.}

\begin{document}

%% LaTeX will automatically break titles if they run longer than
%% one line. However, you may use \\ to force a line break if
%% you desire.

\title{Thermal to Nonthermal Energy
   Partition at the Early Rise Phase of Solar Flares}

%% Use \author, \affil, and the \and command to format
%% author and affiliation information.
%% Note that \email has replaced the old \authoremail command
%% from AASTeX v4.0. You can use \email to mark an email address
%% anywhere in the paper, not just in the front matter.
%% As in the title, use \\ to force line breaks.

\author{Alexander A. Altyntsev\altaffilmark{1}, Gregory D. Fleishman\altaffilmark{2,3}, Sergey V. Lesovoi\altaffilmark{1}, Natalia S. Meshalkina\altaffilmark{1}}
\email{altyntsev@iszf.irk.ru} \altaffiltext{1}{Institute of
Solar-Terrestrial Physics,
 Siberian Branch of the Russian Academy of
           Sciences,
           P.O. Box 4026, Irkutsk 33, 664033, Russia}
 \altaffiltext{2}{Center For Solar-Terrestrial
Research, New Jersey Institute of Technology, Newark, NJ 07102}
\altaffiltext{3}{Ioffe Physico-Technical Institute, St. Petersburg
194021, Russia}

%% Mark off your abstract in the ``abstract'' environment. In the manuscript
%% style, abstract will output a Received/Accepted line after the
%% title and affiliation information. No date will appear since the author
%% does not have this information. The dates will be filled in by the
%% editorial office after submission.

\begin{abstract}
In some flares the thermal component appears much earlier than
the nonthermal component in X-ray range.
Using sensitive microwave observations we revisit this finding made
by Battaglia et al. (2009) based on thorough analysis of %X-ray
RHESSI data.  We have found that
nonthermal microwave emission produced by accelerated electrons
with energy of at least several hundred keV, appears as early as
the thermal soft X-ray emission indicative that the electron
acceleration takes place at the very early flare phase. The
non-detection of the hard X-rays at that early stage of the flares
is, thus, an artifact of a limited RHESSI sensitivity. In all
considered events, the microwave emission intensity increases at
the early flare phase. We found that either thermal or nonthermal
gyrosynchrotron emission can dominate the low-frequency (optically
thick) part of the microwave spectrum below the spectral peak
occurring at 3-10 GHz. In contrast, the high-frequency \textit{optically thin} part of
the spectrum is always formed by the \textit{nonthermal}, accelerated
electron component, whose power-law energy spectrum can extend up to
a few MeV at this early flare stage. This means that even though
the total number of accelerated electrons is small at this stage,
their nonthermal spectrum is fully developed. This
implies that an acceleration process of available seed particles is
fully operational. While, creation of this seed population (the
process commonly called `injection' of the particles from the
thermal pool into acceleration process) has a rather low
efficiency at this stage, although, the plasma heating efficiency
is high. This imbalance between the heating and acceleration (in
favor of the heating) is difficult to reconcile within most of
available flare energization models. Being reminiscent of
the trade off between the Joule heating and runaway electron
acceleration, it puts  additional constraints on the
electron injection into the acceleration process. As a byproduct
of this study, we demonstrate that for those cases when the optically thick part of
the radio spectrum is dominated by the thermal  contribution, the microwave spectral data
yields reliable estimates of the magnetic field
and source area at the early flare phase.
\end{abstract}

%% Keywords should appear after the \end{abstract} command. The uncommented
%% example has been keyed in ApJ style. See the instructions to authors
%% for the journal to which you are submitting your paper to determine
%% what keyword punctuation is appropriate.


%% From the front matter, we move on to the body of the paper.
%% In the first two sections, notice the use of the natbib \citep
%% and \citet commands to identify citations.  The citations are
%% tied to the reference list via symbolic KEYs. The KEY corresponds
%% to the KEY in the \bibitem in the reference list below. We have
%% chosen the first three characters of the first author's name plus
%% the last two numeral of the year of publication as our KEY for
%% each reference.

%% Authors who wish to have the most important objects in their paper
%% linked in the electronic edition to a data center may do so by tagging
%% their objects with \objectname{} or \object{}.  Each macro takes the
%% object name as its required argument. The optional, square-bracket
%% argument should be used in cases where the data center identification
%% differs from what is to be printed in the paper.  The text appearing
%% in curly braces is what will appear in print in the published paper.
%% If the object name is recognized by the data centers, it will be linked
%% in the electronic edition to the object data available at the data centers
%%
%% Note that for sources with brackets in their names, e.g. [WEG2004] 14h-090,
%% the brackets must be escaped with backslashes when used in the first
%% square-bracket argument, for instance, \object[\[WEG2004\] 14h-090]{90}).
%%  Otherwise, LaTeX will issue an error.

\section{Introduction}

The most prominent flare manifestations are explosive plasma
heating and particle acceleration. The balance between thermal and
nonthermal energy content varies during the flare and from one
flare to another. The particle acceleration is widely believed to
dominate at the impulsive phase; these electrons, after being
accelerated, propagate along closed magnetic field lines from
coronal acceleration sites towards denser low corona and
chromosphere. A direct diagnostics of the nonthermal electrons
comes from hard X-ray observations of the bremsstrahlung as the
flare-accelerated electrons collide with ions. The bremsstrahlung
intensity is proportional to the ambient density; so the bulk of
the hard X-ray (HXR) emission comes typically from the
chromospheric footpoints  where the precipitating
flare-accelerated electrons lose most of their energy. In
contrast, coronal HXR emission from the very acceleration sites is
difficult to detect except  {cases of acceleration in very
dense flaring loops \citep{bastian2007, Xu_etal_2008}. However, it
is difficult here to isolate the directly accelerated component
because of essential \textit{in-situ} Coulomb losses, whose effect
depends on both transport regime and particle energy. Another
class of events in which the coronal HXR emission is easier to
detect is  the partially occulted flares} where the intense
footpoints are not visible \citep[see][]{masuda1995,
Wang_etal_1995, Krucker_etal_2010}, and a review by
\citet{krucker2008}). Analysis of such cases, however, suffers
from the fact that the full flare energy balance is unavailable
since the footpoint information is missing. Fortunately, the
situation is greatly improved if   microwave observations are used
{to complement the X-ray data}. The radio emissions
produced directly at the coronal acceleration site can be firmly
separated from other competing radio components, in favorable
cases, allowing us to study the acceleration region directly
\citep{Fl_etal_2011}.

In addition to the main flare phase showing copious electron
acceleration; observations provide strong evidence for a pre-flare
coronal activity unrelated to evaporation driven by precipitating
electrons \citep{harra2001}. Sometimes there is a prominent
coronal soft X-ray (SXR) emission occurring several minutes before
the impulsive phase start \citep{farnik1998}. Further important
knowledge about the pre-flare activity has been obtained from
RHESSI spectral observations at the low-energy range, 3 -- 20 keV.
\citet{krucker2008} called this flare period with a pronounced
gradual increase of the SXR emission occurring minutes before the
detectable onset of the HXR emission the \textit{early flare
phase}. The flares occurred on 2002 July 23 \citep{Lin_etal_2003,
asai2006} and 2003 November 03 \citep{veronig2006} represent vivid
examples of flares with the early phase.

The study of flare energization in the form of plasma heating and
particle acceleration at this early phase is of particular
interest, because it can clarify how the flare energy release is
initiated and what mechanisms are responsible for partition of the
released energy between the thermal and nonthermal components.
{For example, recently \citet{battaglia2009} have analyzed
four flares and have demonstrated that morphology of these events
showed one hot coronal source only, while the footpoints appeared
much later, at the onset of the nonthermal HXR emission.
Remarkably,  the X-ray spectra of these coronal sources at the
early phase were nicely fit as purely thermal.
\citet{battaglia2009} concluded that (i) the early flare phase is
purely thermal; (ii) the detected increase of the thermal plasma
emission measure (EM) requires some chromospheric evaporation, and
(iii) the heat conduction is needed to drive this chromospheric
evaporation.} On the other hand \citet{siarkowski2009} have
studied the early phase in two simple flares, and concluded that
the process of electron acceleration can occur during the early
phase of flares; well before the impulsive stage.

Microwave observations of the 2002 July 23 flare
\citep{asai2006, asai2009} favor the non-thermal alternative
\citep[see, also,][]{white2011}. \citet{holman2003} examined
HXR features of this flare, and reported that the nonthermal
energy, even before the impulsive phase, was quite large. So, there
is yet no clear and consistent picture of how the flare/preflare
energy is released, and in what proportions it is divided between
the thermal and nonthermal components. In particular, if the
plasma is primarily heated by Coulomb losses of accelerated
electrons, or directly by the energy release remains an open,
fundamental problem \citep[see, e.g.][]{petrosian2004, liu2009}.

In this study we address this fundamental problem using microwave
data (from various instruments), which is more sensitive to the
nonthermal electron component than the HXR data. Along with other
available microwave data, we use an essentially new input for the
early phase study: the radio imaging at 5.7~GHz, which is indeed
of  primary importance since this frequency falls into the
spectral peak region of the radio emission from the coronal sources at
the early phase. Note, that the first observation of the 5.7 GHz
early coronal source was performed back in 1992, well before the
RHESSI era during X9 limb flare on 1992 November 02
\citep{altyntsev99}. The brightness center at 5.7~GHz was 10~Mm
higher than the 17 GHz footpoint source.

{Observationally, this study can be facilitated by the fact
that the preflare coronal sources are often reasonably compact and
weak. So the role of source inhomogeneity and contribution from
competing emission mechanisms, otherwise essential, can be
relatively minor in this case. Our event list consists of two (of
four) events from \citet{battaglia2009} list complemented by three
more events suitable for the analysis. No appropriate microwave
data is available for two other events from \citet{battaglia2009}
list. For the modeling of the microwave spectra we take into
account both thermal and nonthermal electron components
\citep{benka1992, fleishman2010}.}

In all the cases, we firmly detected a nonthermal electron
population as early as the thermal emission appeared. The energy
content of the nonthermal electron population remained small and
inessential for the plasma heating, indicative that the thermal
plasma is heated by the flare energy release directly. In three
(of five) cases, the low-frequency optically thick part of the
microwave spectrum was entirely formed by SXR-producing thermal
plasma, which offers a precise diagnostics of the source effective
area and magnetic field. The source area (and thus, presumably the
source volume) was found to grow in all these three events
indicative that no chromospheric evaporation was needed to account
for the EM increase. The observed EM increase is well consistent
with a growing source but without any change of the thermal electron density.
Although the overall nonthermal electron energy (and number
density) content was small in all the cases, the acceleration
mechanism was fully operational, and powerful enough to produce a
well-developed nonthermal power--law tails of the accelerated
electron spectra, up to a few MeV, typical also for the impulsive
flare phase. This implies that the key distinction between the
`normal' flare and the early flare phase is in remarkably
different efficiency of forming a seed suprathermal electron
population via their extraction from the thermal pool. As soon as
this seed population is formed, it is efficiently transformed to a
power--law nonthermal distribution by the main acceleration
process. Stated another way, the processes of electron injection
from the thermal pool and their further acceleration are driven by
distinctly different  physical processes. The former process seems
to be responsible for the energy partition between the thermal and
nonthermal components, in the way that the stronger the plasma
heating the weaker the injection, and vice versa.

\section{Instrumentation and observations}

The events under study were observed by a number of radio
instruments: The Nobeyama Radio Polarimeters
\citep[NoRP,][]{torii1979, nakajima1985} record  total and
circularly polarized intensities (Stokes parameters I and V) at 1,
2, 3.75, 9.4, 17, 35, and 80 GHz (Stokes I only). We also used the
Radio Solar Telescope Network (RSTN) data. One second temporal
resolution data were taken at 8 selected frequencies (0.245, 0.41,
0.61, 1.415, 2.695, 4.995, 8.8, 15.4~GHz) from Learmonth station,
Australia.

\begin{figure*}[!t]
\epsscale{.80}
\plottwo{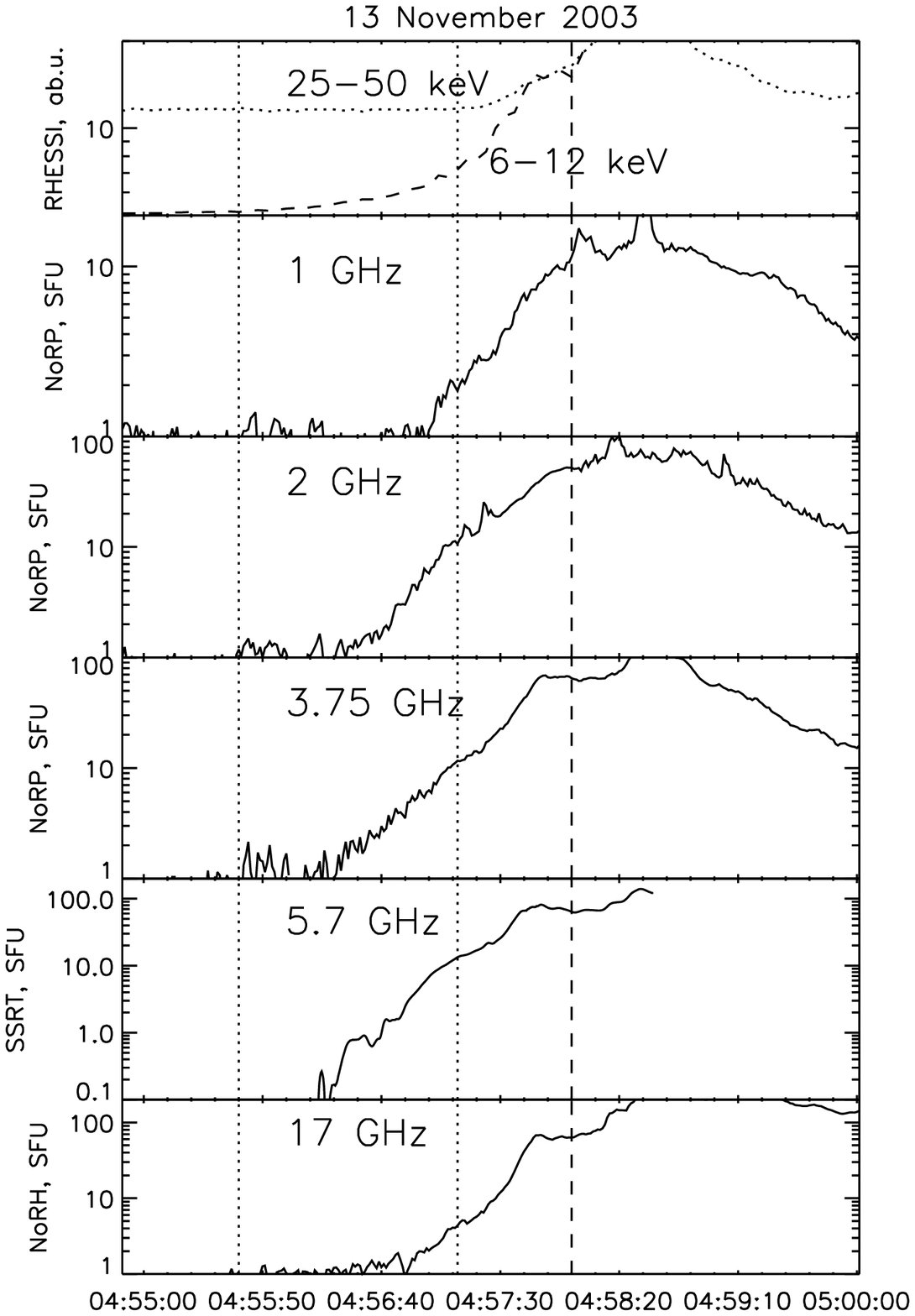}{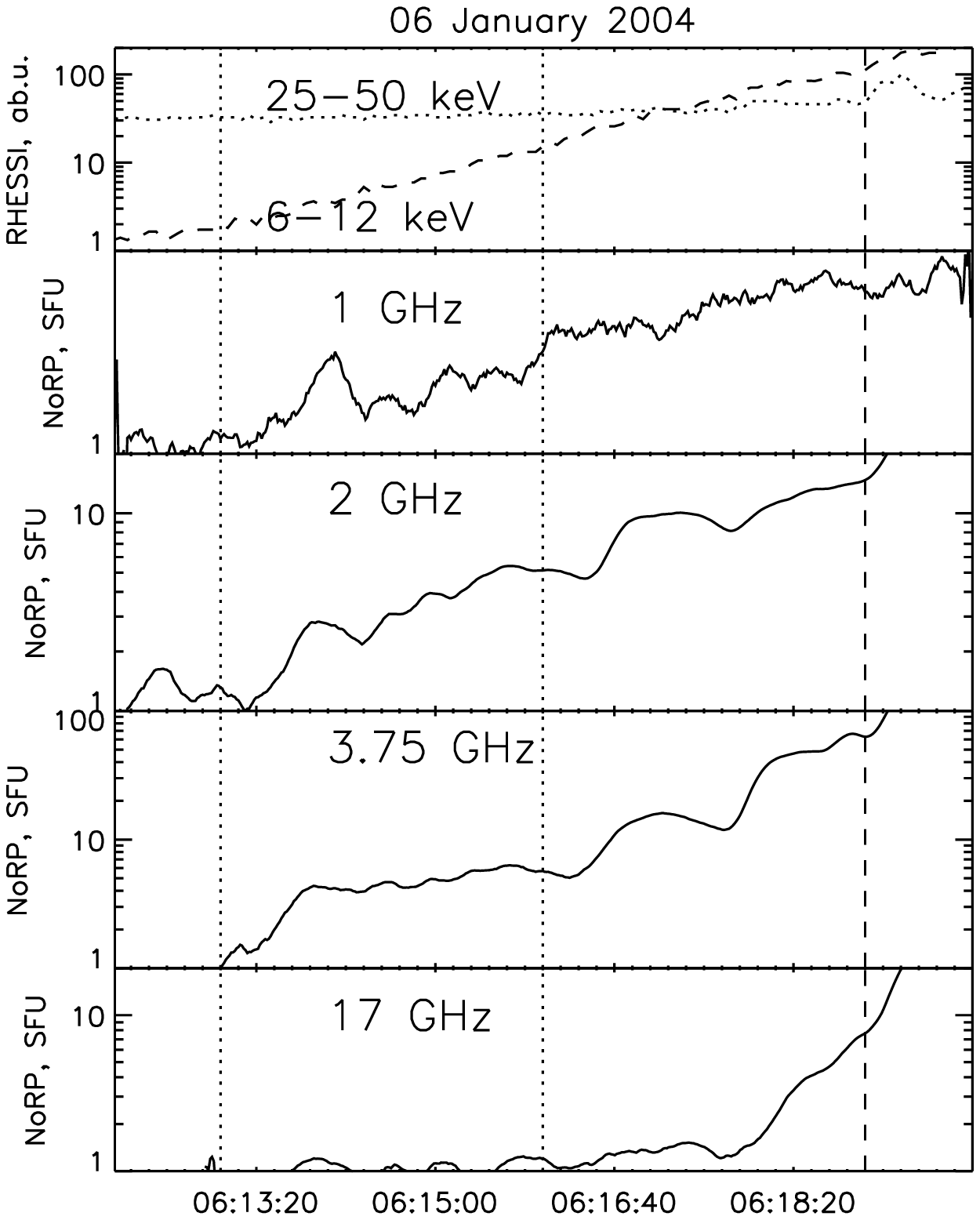}
%\centerline{\hspace*{0.015\textwidth}
 %       \includegraphics[width=0.455\textwidth,clip=]{Fig1a.eps}
  %      \includegraphics[width=0.470\textwidth,clip=]{Fig1b.eps}
   %           }
    % \vspace{-0.35\textwidth}   % Shift close to the panel top
     %     \centerline{\large      % Includes the labels (here needs the color
      %                          %   package, see beginning of this file)
%      % \hspace{0.13 \textwidth}  {CEB}
        % \hfill}
     %\vspace{0.33\textwidth}
\caption{Top: RHESSI corrected count rate light curves in 6--12
keV (dashed), and 25--50 keV (dotted), during the early phase of the
two events; same background is used as in \citet{battaglia2009}. Below: Microwave profiles recorded by NoRP, NoRH, and
SSRT. The dotted vertical lines indicate the start and end of  the early phase, the dashed lines show the time of
the \textbf{X-ray} emission appearance at footpoints.}
\label{Fig1}
\end{figure*}
%\clearpage

The Nobeyama Radioheliograph \citep[NoRH,][]{nakajima1994} obtains
images of the Sun at 17 GHz (Stokes I and V) and 34 GHz (Stokes
I). The NoRH angular resolution is around 10 arcseconds at 17 GHz,
and 6 arcseconds at 34 GHz. Siberian Solar Radio Telescope
\citep[SSRT;][]{smolkov1986, grechnev2003} produces microwave
images at 5.7 GHz. In some events we also used one-dimensional
brightness distributions measured with NS and EW arms of the SSRT.
To analyze the most recent event in our event list (2011 February 15) in greater
detail, we take advantage of more complete microwave observations
obtained from new instruments developed by the SSRT team: the
spectropolarimeter recording the total power data at 2.3, 3.0,
4.0, 5.0, 6.0, 7.0 and 8.0 GHz (SP 2--9), and the 10-antenna
prototype of the multifrequency heliograph (mSSRT) observing at
5.1, 5.5, 6.3, 6.9 and 7.6 GHz \citep{lesovoy2009, lesovoi2012}.

We also use magnetograms from the Michelson Doppler Imager
\citep[MDI,][]{scherrer1995}, and EUV images from the
Extreme-Ultraviolet Imaging Telescope
\citep[EIT,][]{delaboudiniere} on SOHO; as well as X-ray data from
the Ramaty High-Energy Solar Spectroscopic Imager
{\citep[RHESSI,][]{lin02}} and Fermi Gamma-ray Space
Telescope \citep{Meegan2009}.

\textbf{Flares on 2003 November 13 (N01E80) and 2004 January 6
(N05E89)}. These two flares from the \citet{battaglia2009} event
list occurred during the SSRT observing time. The RHESSI corrected
count rate light curves in 6--12 keV and 25--50 keV are shown in
Figure \ref{Fig1}. The early phase is marked by the dotted
vertical lines (04:55:40--04:57:12 UT and 06:13--06:16 UT) and
corresponds to the time interval of gradual increase of the 6--12~keV light curve before appearance of the 25-50 keV emission
enhancement \citep[see Figure 1 from][]{battaglia2009}. The
microwave light curves recorded with the NoRP, NoRH and the SSRT
are also shown. Note that the microwave emission at frequencies
below 10 GHz appeared well before the impulsive phase.

\begin{figure}
 \centering
\includegraphics[width=0.99\columnwidth]{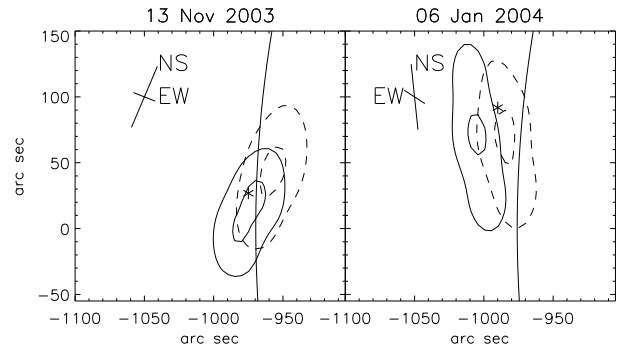}
\caption{Contours: microwave sources before flares (dashed) and at
the end of the early phase (solid). Contour levels are $9 \times
10^{5}$ K and $1.6 \times 10^{6}$ K for 13 November, and $3.4
\times 10^{5}$ K and $6.2 \times 10^{5}$K for 6 January.
 {The half widths of the SSRT beams, whose sizes and directions are dependent on the local time of observation, are shown by crossing
bars} in the top left corner. The half widths of the SSRT beam:
$17\times50$ arcseconds (13 Nov) and $18\times50$ arcseconds (6
Jan).  {Line directions correspond to NS/EW scanning
directions at times of the observations.} The asterisks show
centroid positions of the coronal X-ray sources from
\citet{battaglia2009}.} \label{Fig2}
\end{figure}

Images of the microwave sources at 5.7 GHz are shown in
Figure~\ref{Fig2} before the flare and at the early flare phase.
{Strong background microwave (presumably, gyroresonance)
sources observed before the flares were situated close to the limb
above AR 10537 and AR 10591, respectively} (the dashed contours in
Figure~\ref{Fig2}). The corresponding microwave sources at the end
of the early phase (the solid contours obtained after the
background subtraction) were located above the limb at distances
exceeding the half widths of the SSRT beam. \citet{battaglia2009}
have studied the X-ray sources at this early phase in detail.
Centroids of the X-ray loop top sources shown by the asterisks in
Figure \ref{Fig2} are located clearly above the limb, implying the
corresponding coronal loop top sources. The apparent sizes of the
coronal microwave sources are larger than the corresponding X-ray
sources because of relatively low spatial resolution of the SSRT,
shown by the NS/EW crosses. The X-ray source volumes determined
from the SXR imaging were $4.8 \times 10^{26} $cm$^{3}$ and $2.8
\times 10^{27}$cm$^{3}$ {by the end of the early stage},
when their apparent sizes were $10.8''$ and $19.4''$, accordingly.
{During the early stage the source size remained roughly
constant during 13 November early phase event, while  the area of
the 06 January source increased, roughly, by a factor of three}.
The emission measures and temperatures determined using the RHESSI
observations were gradually growing up to $1.7 \times 10^{46}
$cm$^{-3}$ and 25 MK for 13 November event, and $1.1 \times
10^{46} $cm$^{-3}$ and 35 MK for 06 January event.

\begin{figure}
\epsscale{1.2}
 \includegraphics[width=0.49\columnwidth]{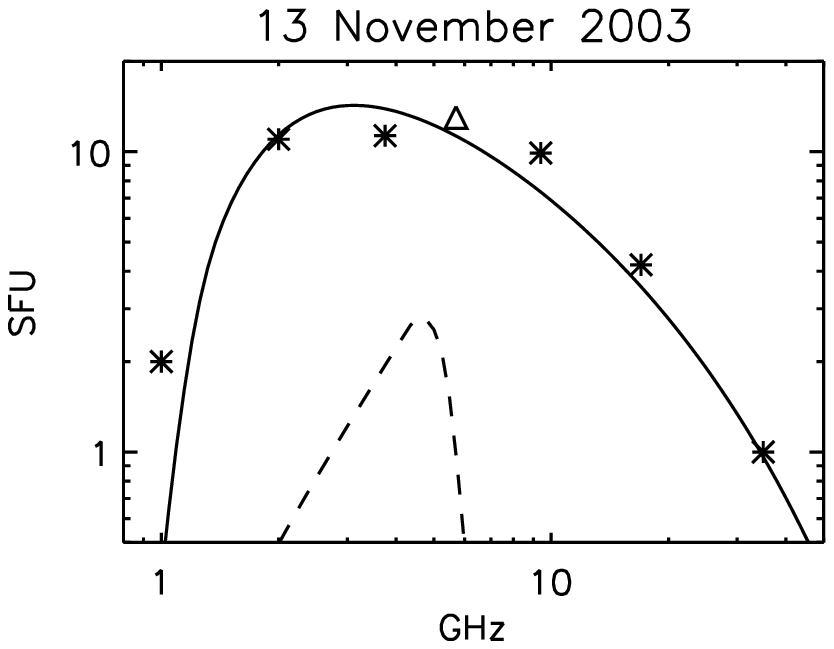}
 \includegraphics[width=0.49\columnwidth]{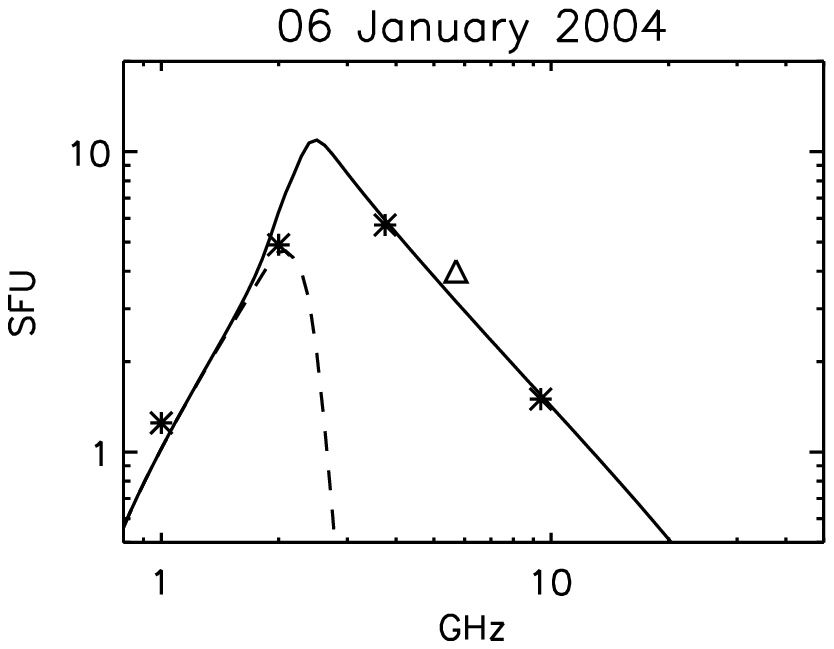}
\caption{Microwave spectra at the end of the early phase and
results of fitting by gyrosynchrotron emission generated by
thermal and nonthermal electron distribution. Dashed curves show purely thermal
gyrosynchrotron emission. Observations from
NoRP (asterisks) and SSRT (triangles).} \label{Fig3}
\end{figure}

The microwave spectra observed at the end of the early phase are
shown in Figure~\ref{Fig3}. The turn-over frequencies in both
spectra were below the SSRT operating frequency implying that SSRT
observed in the optically thin part of the microwave spectrum;
{i.e., the \textit{optically thin} radio emission was produced from
a \emph{coronal}, not the footpoint, part of the source.}
%\textbf{so there were no simultaneous HXR footpoint sources in
%relatively strong magnetic field regions in the photosphere}.
Thus, we can confidently conclude that the total power spectra
obtained at other frequencies without spatial resolution pertain
to coronal sources corresponding to the SSRT sources observed
above the limb.

\textbf{X4.8 Flare on 2002 July 23 (S13E72).} The coronal X-ray
source appeared at the initial stage of this flare. It is the
first proposed example of the early phase sources
\citep{krucker2008}. The early phase (00:18 -- 00:26 UT) continued
until the HXR emission of the coronal source remained dominating.
The X-ray observations at this phase were also analyzed by
\citet{Lin_etal_2003}, \citet{holman2003}, and \citet{caspi2010b}.
The coronal source displayed a gradual increase in both SXR and
HXR emission below 60 keV. The RHESSI signal at higher energies
appeared after 00:26 UT only. \citet{caspi2010b} have found three
coronal electron populations: nonthermal, super--hot thermal, and
hot thermal. Temperatures of the thermal electrons  {were
determined at 00:20 UT and then were roughly} constant and equal
to $\sim $30~MK and $\sim$ 20~MK, respectively \citep[see Figure 1
in][]{caspi2010b}. The plasma density  {of the hot
component} was estimated to be $1.7 \times 10^{10}$cm$^{-3}$
 {\citep[Fig.4.10 from][]{caspi2010a}}. The HXR observations
show that these plasma parameters did not change %considerably until 00:24:00 UT.
 {noticeably  from 00:20 UT to 00:24 UT. However,  the
source volume was increasing from $3.7 \times 10^{26}$ cm$^{-3}$
up to $7 \times 10^{26}$ cm$^{-3}$ during this same time
interval.}

\begin{figure}
\epsscale{.990} \plotone{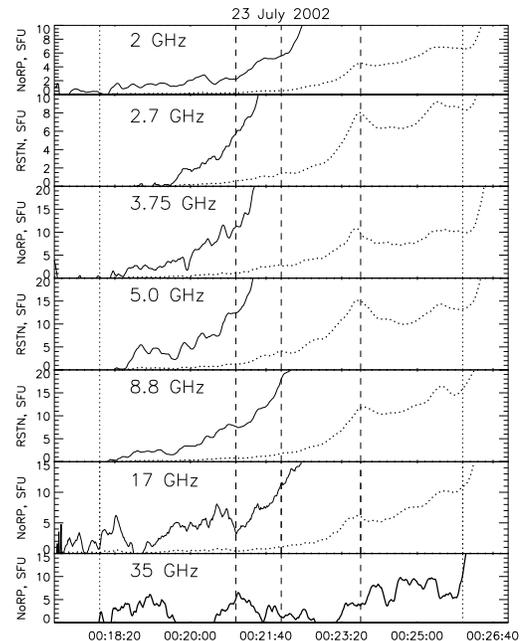}
\caption{Profiles of the integral microwave fluxes, recorded with
the NoRP and RSTN. The dotted curves show  the same fluxes divided
by ten. The early phase bounds (00:18 -- 00:26 UT) are marked by
the dotted lines.} \label{Fig4}
\end{figure}

\begin{figure}
\epsscale{1.2} \plotone{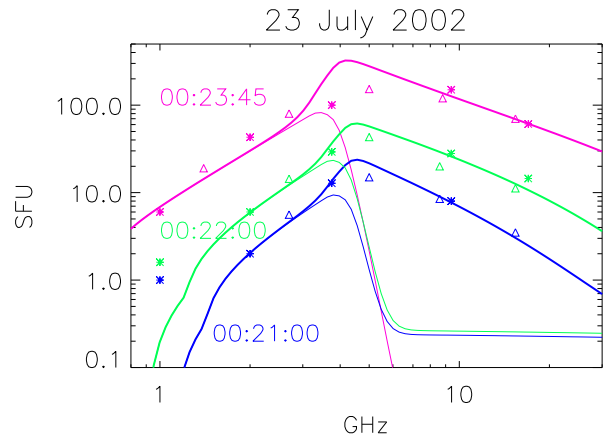} \caption{Spectra recorded with
NoRP (asterisks) and RSTN (triangles) at the moments marked in
Fig.\ref{Fig4} by vertical dashed lines. The curves present
results of spectral fitting by the TNT (thick) and the THM (thin)
codes.} \label{Fig5}
\end{figure}

The microwave light curves (Figure \ref{Fig4}) grew gradually
during the early phase, and sharply afterward. The SSRT
observations have began at the impulsive phase only,  after 00:33
UT; although radio  emission at frequencies up to 17 GHz  appeared
during the early {phase. At} the end of this phase the NoRP signal
at 35 GHz appeared. A more sensitive Nobeyama Radioheliograph
recorded the flare emission at 17 and 34 GHz well before that
time. After the 34 GHz emission appearance at 00:22 UT the main
microwave source was located slightly above the SXR flare loop
\citep{white2003, asai2006, asai2009}. The footpoint emission also
appeared and  became essential after 00:22:30 UT \citep{asai2006}.
The spectrum evolution at the early phase is shown in Figure
\ref{Fig5}.  Good correspondence between the NoRP and the RSTN
data gives an idea on the measurement accuracy.

\textbf{X7.1 Flare on 2005 January 20 (S06W90).} Many studies were
devoted to this extreme event, but they are mainly concerned the
origin of accelerated protons, relativistic electrons, and the
cosmic ray aspect \citep[see, e.g.,][]{grechnev2008}. A gradual
increase of SXR flux started at 06:00~UT,  according to GOES-10 and
GOES-12. At least half an hour before the flare maximum at 06:46
UT, RHESSI detected the blob at loop top\footnote{
http://svs.gsfc.nasa.gov/vis/a000000/a003100/a003162/.}.

\begin{figure}
 \centering
     \includegraphics[width=0.99\columnwidth]{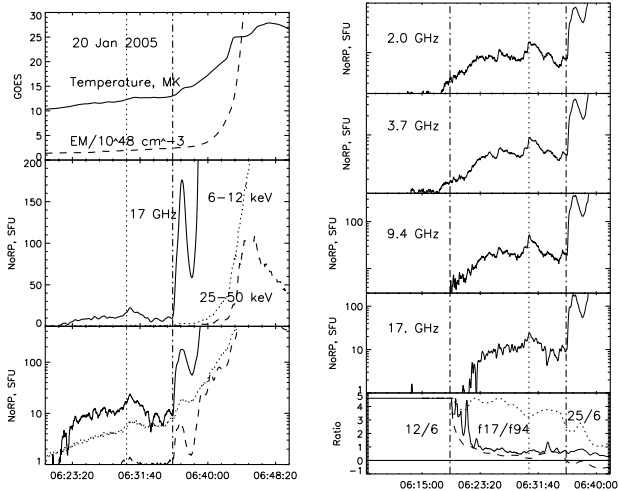}
\caption{a) X-ray and microwave profiles in the early phase before
the flash phase started at 06:35:40 UT (marked by dash-dotted line).
Dotted line illustrates the end of the Nobeyama observations. Top: GOES
temperature and emission measure.  {Background microwave and
X-ray fluxes are subtracted.} Bottom: NoRP and RHESSI light curves
(in arbitrary units) in linear and logarithm scales. b) NoRP
microwave profiles in the early phase. Bottom panel: natural
logarithms of flux ratios in microwaves and X-rays. Vertical lines
correspond to 06:20, 06:30:20, and 06:35:40 UT.} \label{Fig6}
\end{figure}

The early phase lasted until 06:35:40 UT, where the temperature
determined from GOES 10 data increases gradually from 7 to 13 MK
without a noticeable increase of the emission measure
 {(Figure \ref{Fig6})}. The HXR emission flux exceeds the
background level at energies below 40 keV, where it can be fitted
by a power--law with index 7.3 at 06:30 UT.

\begin{figure}
 \centering
\includegraphics[width=0.99\columnwidth]{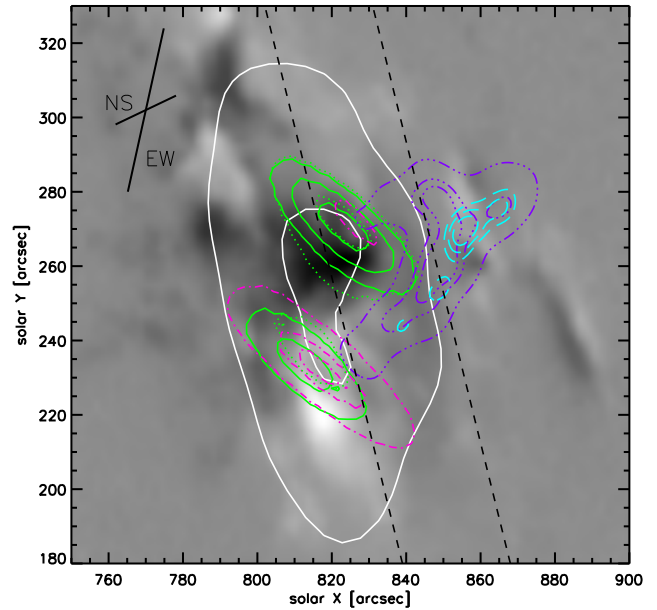}
\caption{Flare structure in the early phase. MDI magnetogram
(background, 05:46:30 UT); 3 -- 12 keV RHESSI images (violet
dash-dot-dotted contours at 0.3, 0.65, 0.8  levels, 06:41:00 --
06:41:30 UT); 12 -- 25 keV(turquoise dashed at 0.5, 0.65, 0.8
levels, 06:29:00 -- 06:32:00 UT); 5.7 GHz SSRT image (white solid
contours at 0.5, 0.9 levels, 06:33:26 UT); 17 GHz NoRH image in
intensity (green continuous contours at 0.5, 0.7, 0.9 levels,
06:30 UT) and in the left circular polarization (green dotted at
0.9, 0.5 levels, 06:30 UT); 34 GHz NoRH image in intensity (pink
dash-dotted contours at 0.35, 0.7, 0.9 level, 06:30 UT). The black
dotted lines show the directions of the SSRT EW-array fan.
 {The half widths of the SSRT beams are shown by crossing
bars in the top left corner: $18\times45''$.}}
\label{Fig7}
\end{figure}

First signatures of the nonthermal electrons appeared in
microwaves after 06:20 UT. The radio fluxes at 2 -- 17 GHz were
synchronously increasing during fifteen minutes and reached a
level of 100 SFU at 3 -- 5 GHz.  The microwave profiles are
similar to X-ray profiles at energies below 25 keV (see Figure
\ref{Fig6}b, bottom). The degree of polarization
 of the total power emission was 10--20\% LCP and almost
did not change during the early phase at frequencies above 3.75
GHz. At 2 GHz the degree of polarization reached 30\% RCP at some
time frames. The similarity of the microwave and X-ray light
curves favors a common origin of radio-- and X-ray-emitting
electrons. Sequence of the RHESSI images in the 3 -- 12 keV
channels shows that the coronal source was located within the
flare loop visible later in the impulsive stage (Figure
\ref{Fig7}), and it was purely coronal until 06:38 UT.  {The
source volume  was increasing during the first ten minutes of the
early phase.} \citet{grechnev2008} have estimated the loop
parameters determined from EIT/171A and GOES/SXI images at
06:42:20 UT. The loop height was $4.6 \times 10^{9}$~cm, and width
$9 \times 10^{8}$ cm (volume $7.8 \times 10^{27}$ cm$^{3}$).
Plasma density was estimated to be $(1-2) \times 10^{10}
$cm$^{-3}$.

\begin{figure}
 \centering
 \includegraphics[width=0.99\columnwidth]{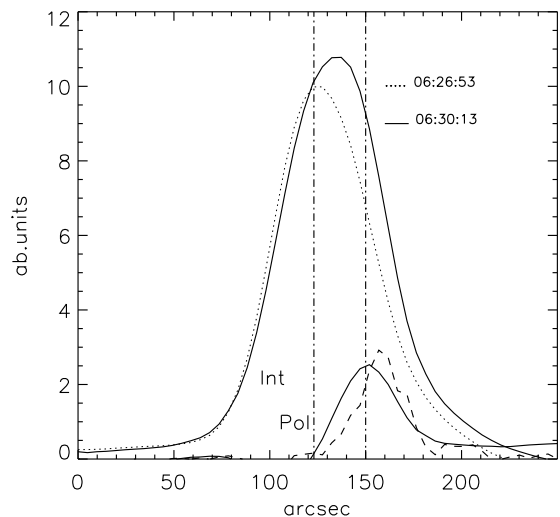}
\caption{One-dimensional distributions of the flare region at two
moments during the early phase (solid and dotted curves). Vertical
lines correspond to the EW fan directions in Fig.\ref{Fig7}.
 {Lower solid and dashed curves show the subtraction of the
two one-dimensional distributions in intensity and polarization, and
correspond to the flare coronal source.} Dashed curve shows
distribution of circular polarization.} \label{Fig8}
\end{figure}

Microwave sources are shown in Figure \ref{Fig7} at the end of the
NoRH observations (06:30 UT).  {The 17 GHz and 34 GHz
emission sources are seen to originate at loop footpoints, located
in the regions with opposite polarities of the photosphere
magnetic field.} A line-of-sight magnetic field of 1200 G
(N--polarity) was measured in the South footpoint, while $-1300$~G
(S polarity) in the North footpoint. At the early phase the flare
emission at 5.7 GHz was relatively weak compared with the
background gyroresonant emission, which does not favor its direct
imaging. However, the coronal source can be revealed by
subtracting the background one-dimensional intensity distribution
(Figure \ref{Fig8}). One-dimensional distributions were recorded
in the direction perpendicular to the fans of the linear arrays
every 0.3 sec. The dotted lines in Figure \ref{Fig7} show the
directions of the EW/SSRT array fan.

\begin{figure}
 \centering
\includegraphics[width=0.99\columnwidth]{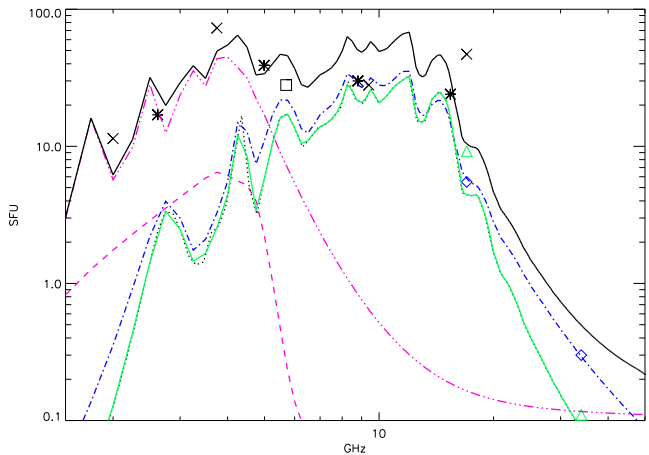}
\caption{ Microwave spectrum at 06:30 UT, recorded by
NoRP(crosses), RSTN(asterisks), SSRT(squares), NoRH in the North
footpont source (triangles), and the South source (diamonds).
 {The THT model spectra were calculated for loop top source
(pink dashdot-dotted solid), the North footpoint source} (green
solid), and the South source (blue dash-dotted). The sum of the
model  {spectra is shown by solid black thick curve. The
dashed pink curve shows the contribution of gyrothermal emission
from the coronal source.}} \label{Fig9}
\end{figure}

\begin{figure}
 \centering
\includegraphics[width=0.99\columnwidth]{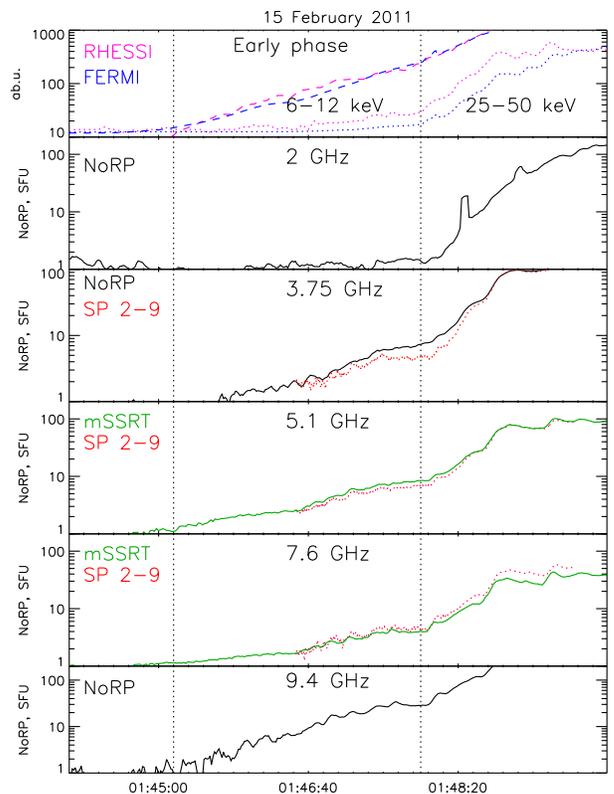}
\caption{Top panel: RHESSI and Fermi X-ray signals at energies 6--12 keV (pink
dashed) and 25--50 keV(blue dotted). Background microwave and
X-ray fluxes are subtracted. Below: temporal profiles recorded
with spectropolarimeters NoRP, SP~2--9, and mSSRT. The vertical
lines bound the early phase.} \label{Fig10}
\end{figure}

Figure \ref{Fig8} presents one-dimensional distributions of flare
region before and during the brightening at 06:30 UT. The
subtraction of the former from the latter reveals the early flare
phase source. The flare source was displaced to the West from the
brightness center of the background source, so its location is
likely to correspond to the looptop region in Figure \ref{Fig8}.
The FWHM size of the source is about $30''$, while the flux
level of 28~sfu is about the total flare flux at this frequency. The
degree of polarization reached 20\% RCP; the brightness center of
the polarized emission was displaced by $10''$ from the
intensity brightness center to the North-West. Thus, the
bulk of the 5.7 GHz flare emission was produced from {the}
coronal source.

The microwave spectrum  {corresponding to the emission
enhancement at 06:30 UT} is shown in Figure \ref{Fig9}. The data
at frequencies 2, 3.75, 9.4 GHz were recorded with the Nobeyama
spectropolarimeter, at frequencies 1.4, 2.65, 4.99, and 8.8 GHz
with the Learmonth Solar Observatory (USAF RSTN Network)
spectrometer. The SSRT flux at 5.7 GHz was determined from
 {sequence of the one-dimensional EW distributions.} It is
reasonable to suppose that most of the emission below 6 GHz (see
the radio spectrum in Figure \ref{Fig9}) originates from the
coronal source. The microwave fluxes at %frequencies
17 and 34 GHz,
dominated by the footpoint sources (see Figure \ref{Fig7}) were
obtained using the NoRH observations and are shown by  triangles and diamonds in Figure \ref{Fig9}.
Assuming a power-law spectrum ($F \sim f^{-\gamma}$) at the range from 17 GHz to 34 GHz,
we can estimate $\gamma_{North}=6.4$ at the
North footpoint, and $\gamma_{South}= 4.2$ at the South one.

\begin{figure}
 \centering
\includegraphics[width=0.9\columnwidth]{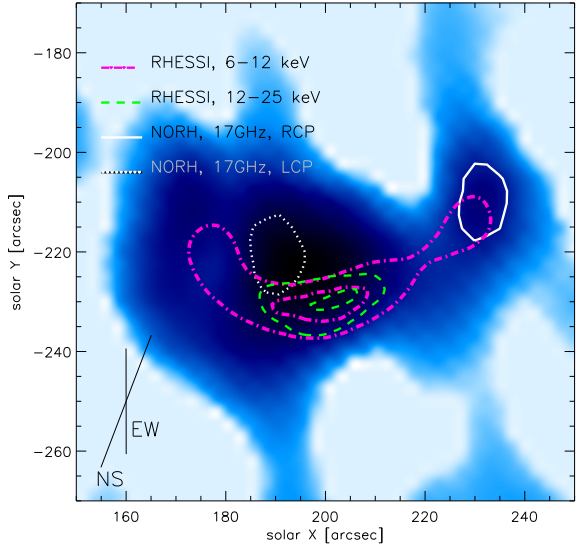}
\caption{Flare on 2011 February 15 at 01:47:55. Microwave emission
at 17 GHz: brightness temperature in intensity (background), in
circular polarization - white contours (right-handed - solid and
left-hand -- dotted). The HXR emission is shown in the 6 -- 12 keV
by the pink dash-dotted contours, and in the 12 -- 25 keV by the
green dashed contour. Axes are scaled in arcseconds. The half
widths of the SSRT beams are shown by crossing bars in the bottom
left corner: $19\times25$ arcseconds.} \label{Fig11}
\end{figure}

\textbf{X2.2 Flare on 2011 February 15 \textbf{(N21W21)}}. The
flare on 2011 February 15 was the first X-class flare in the
current solar activity cycle. Not surprisingly, it has already
been analyzed in a number of papers. This flare occurred in AR
11158 in the central area of the solar disk.  Given that the above
study of the limb early flare phases have established that the
main HXR ($\leq$ 50 keV) and microwave emissions ($\leq$ 10 GHz)
originate from compact coronal sources, it is reasonable to adopt
the same morphology for the disk flares as well.

For this flare analysis we take great advantage of new radio
instruments developed by the SSRT team, and observing now at the
SSRT site. In fact, the 2011 February 15 event is the first
large flare recorded with these new SSRT instruments. The HXR
and microwave profiles at the preflare phase are shown in Figure
\ref{Fig10}. Vertical lines bound an interval 01:45:10 -- 01:47:55
UT. At the end of the early phase the estimates of plasma
temperature and emission measure were 14 MK and $1.2 \times
10^{48} $cm$^{-3}$ from the GOES 15 X-ray data, and 34 MK and $3
\times 10^{46} $cm$^{-3}$ from the RHESSI data. The microwave
emission was gradually rising during the early phase. Note that
there is a good match between the light curves measured with
different instruments. The emission at 17 GHz appeared during the
early phase at 01:46:20 UT, while at 34 GHz even after this phase.

The flare spatial relationships at the end of the early phase are
shown in Figure \ref{Fig11}. There was a loop seen in the 6--12
keV emission and in the microwaves. Its footpoints were
{clearly} marked up by the oppositely polarized sources at
17 GHz. The brightness center of the 12 -- 25 keV HXR source
located at the loop top. The mSSRT observations have shown, through
a forward fit of the measured visibilities, that the 4 -- 9 GHz
sources located near the loop top as well and their sizes were
about $20''$  at the early phase.

\begin{figure}
 \centering
\includegraphics[width=0.99\columnwidth]{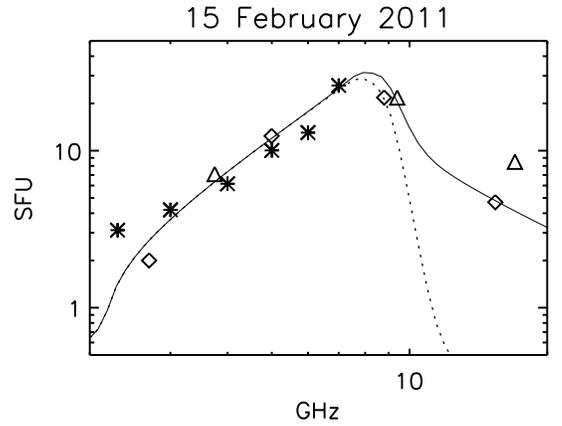}
\caption{Microwave spectrum at the end of the early phase
(01:47:55 UT).} \label{Fig12}
\end{figure}

The microwave spectrum recorded at the end of the early phase with
various instruments is shown in Figure \ref{Fig12}. There is a
good correspondence between the different measurements, except some
mismatch between the NoRH and RSTN fluxes at the frequencies above
15~GHz.

\section{Microwave Spectrum Modeling}

A coronal source (e.g., a magnetic flux tube) can be detected via
microwave observations in a few distinct cases: (i) via the
thermal free-free emission (absorption) in case of high plasma
density (large EM); (ii) via the thermal
gyroresonant/gyrosynchrotron emission (GS) in case of high
magnetic field and significant plasma heating; and (iii) via
various nonthermal emission processes, of which we only consider
below the incoherent gyrosynchrotron emission mechanism.\footnote{Apparently, a combination of the mentioned mechanisms is also
likely \citep[see, e.g.,][]{bastian2007}.} This means that having
enhanced microwave emission does not necessarily imply that this
emission is produced by a \textit{nonthermal} electron population. In
practice, the presence of the nonthermal component can be
confirmed by microwave spectral modeling or forward fit, because
the gyrosynchrotron spectra produced by thermal or nonthermal
electrons are distinctly different. Given that the early flare
phase manifests itself as purely thermal in X-ray range
\citep{battaglia2009} we checked if the microwave spectrum can be
consistently described as being produced by the thermal component
alone, and found that this purely thermal model fails in all the
cases. This implies that a nonthermal electron population is
always present.

Therefore, it is reasonable to model the observed microwave
spectra by gyrosynchrotron emission generated by electrons, with a
so-called thermal/nonthermal distribution (TNT) proposed by
\citet{benka1992}. They demonstrated that the thermal plasma
component can dominate the low-frequency optically thick part of
the microwave spectrum. The TNT distribution
\begin{equation}
f(E)dE=dE\left\{\begin{array}{l} \displaystyle
u_{THM}(E),~\textrm{for}~E < E_{cr},\\[10pt]
A_{nth}E^{-\delta},~\textrm{for}~E_{cr}\leq E < E_{\max}.\\[10pt]
\end{array}\right.
\label{mu_dist}
\end{equation}
ensures a smooth transition from the thermal to nonthermal
distribution at a matching energy $E_{cr}$. In the above
expression, $u_{THM}(E)$ is the Maxwellian distribution function,
and $A_{nth} = u_{THM}(E_{cr})E_{cr}^{\delta}$, to make the
function continuous at the point $E_{cr}$.  {In what follows
we refer a nonthermal component to as all particles between
$E_{cr}$ and $E_{\max}$.}

In practice we employed fast gyrosynchrotron codes recently
developed by \citet{fleishman2010}, which, by default, account for the
corresponding free-free contribution along with the GS one. Given
that available data is limited we are forced to adopt a few
simplifications for our spectral modeling. In particular, we adopt
the isotropic electron distribution in all the cases, and uniform
source model in most cases (unless a different is explicitly
stated). Thus, there are overall nine free parameters for the
spectral modeling: $T$ and $n_{th}$, the temperature and the density
of the background thermal plasma; matching energy $E_{cr}$,
power-law index $\delta$,  the maximum nonthermal energy $E_{\max}$,
%which characterize the number density, hardness, and energy range of the nonthermal electrons;
the magnetic field strength $B$, the
source depth $D$, and the area $A$; which characterize the
emitting source, and the viewing angle $\theta$ relative to the
magnetic field direction.

We start from two events from the \citet{battaglia2009} list,
whose morphology was studied in detail, so some guess parameters
such as the source area $A$, viewing angle, density and
temperature of the background plasma can be taken from the X-ray
observations \citep{battaglia2009}. While the depths of the
microwave sources were estimated as $\sqrt A$, uniform source
model is expected to be valid for the compact looptop sources in
these two flares.  {These input  parameters obtained from
the X-ray observations and used for the microwave modeling are
given by underlined italic characters in Table 1.}

 {A classical GS spectrum displays a main spectral peak
formed by the GS self-absorption at a frequency determined by the
condition $\tau(f)\sim1$, where $\tau(f)$ is the
frequency-dependent optical depth of the GS source and can also
contain a number of secondary (harmonic) peaks at small integer
multiples of the gyrofrequency \citep[see, e.g., examples given in
Figures by][and references therein]{benka1992, fleishman2010}.}
The fast GS code is capable of reproducing the GS harmonic
structure at relatively low frequencies; however, frequency gaps
in the observations are too large to expect any reliable detection
of the harmonic structure. In addition, the harmonic structure is
expected to be smoothed out by a relatively modest non-uniformity
of the coronal source. Thus, we employ a \textit{continuous}
version of the GS code \citep[see][for the code nomenclature
definition]{fleishman2010}, which ensures high overall accuracy,
while smoothies out the low-frequency harmonic structure.

The results of the TNT fitting are shown by the thick curves in
Figure \ref{Fig3}, while the dashed curves show the calculated
spectra for the corresponding purely thermal electron
distribution. Remarkably, the best fit spectra for these two
events are distinctly different from each other. Indeed, for the
13 November 2003 event, the thermal contribution is negligible even
if we adopt a relatively high magnetic field, 300~G, (a
corresponding model spectrum is shown in Figure \ref{Fig3}, left,
dashed curve). This is in fact larger than needed to fit the
high-frequency optically thin part of the microwave spectrum. The
entire spectrum can be nicely fit by a nonthermal GS emission, if
we adopt the source magnetic field to be above 70~G.  {The number
density of nonthermal electrons $n_{nth}$ was calculated using
power law index $\delta$, for energies above $E_{cr}$, and it is
much less than $n_{th}$. Nonthermal energy content was relatively
small also ($W_{nth}/W_{th}\ll 1$). Note that the ratio of the
plasma pressure to the magnetic pressure, commonly called the plasma $\beta$,
is about 0.2 for $B=70$~G.}

In contrast, the thermal GS contribution (THM) entirely dominates
the optically thick part of the microwave spectrum in the 06
January 2004 event; while the high-frequency optically thin part
of the spectrum still has a clearly nonthermal origin. This
thermal part of the GS spectrum is extremely conclusive in terms
of the source parameter diagnostics. Indeed, in the optically
thick regime, the brightness temperature of the thermal emission
equals to the electron kinetic temperature $T$. Thus, the total
power flux level is fully defined by the product $A\times T$ of
the source area by the electron temperature, which immediately
provides the effective source area $A$, given that the electron
temperature is independently known from the SXR spectrum. Given
that the plasma temperature does not change much during the
studied early flare phase, being within 20--30 MK, the optically
thick flux density increase can only be attributed to the
corresponding increase of the effective source area and thus, the
source volume.  {The volume increase can, in principle, be
provided by either the source size or volumetric filling factor
increase, from which we favor the former option given that the SXR
imaging (see the previous section) also reveals the SXR source
size increase with time. } In fact, the entire increase of the
emission measure  observed from the SXR data \citep{battaglia2009}
is accounted by this volume increase, and so no additional hot
material (e.g., from a chromospheric evaporation) is needed. Then,
for the given source size, density, and temperature, the spectrum
peak frequency is solely defined by the source magnetic field.
This offers simple and reliable diagnostics of the coronal source
magnetic field.

The 20 January 2005 event is somewhat different from the two
events considered above, since its imaging data is inconsistent
with the uniform coronal source model showing a strong footpoint
contribution at high frequencies (17 and 34 GHz). This is in fact
expected given relatively strong magnetic field values at the
footpoints. Therefore, we explicitly consider three radio sources (one
coronal and two footpoints), to roughly match the total power
spectrum and also spatially resolved high-frequency data points.

We modeled these two footpoint sources separately to match the
corresponding 17 GHz and 34 GHz spatially resolved data.
 {Recall, the \textit{microwave} spectral indices are
strongly different at the different footpoints, i.e.,
$\gamma_{North}=6.4$ and $\gamma_{South}= 4.2$. The HXR footpoint
spectral indices are also often different from each other, but the
difference is typically within 0.6
\citep{Saint-Hilaire_etal_2008}. In contrast, the difference
between the microwave indices is much larger,
$\gamma_{North}-\gamma_{South}=2.2$. A number of mechanisms has
been proposed to account the HXR spectral index difference
including anisotropic acceleration, different column densities in
the opposite footpoints, differently nonuniform ionization, or
effect of trapping due to magnetic mirroring in an asymmetric
magnetic trap    \citep[see][for greater
detail]{Saint-Hilaire_etal_2008}. Of which only the trapping can
have an effect on the microwave gyrosynchrotron spectrum. However,
formation of a noticeably asymmetric flaring loop is unlikely in
our case, because the magnetic fields at the opposite footpoints
are comparable to each other in absolute value, being $1200$~G and
$-1300$~G, respectively. For the GS microwave spectrum from the
footpoints, a much stronger effect on the spectrum slope is
expected from the pitch-angle anisotropy of the  fast electrons
\citep{Fl_Meln_2003} having almost a pancake distribution at the
mirror points. However, because we do not have a reliable ground
to treat this anisotropy quatitatively, we assume } the isotropic
 {velocity} distribution of fast electrons, and match the
high-frequency data points with $\delta_{South}=4.3$ for the
southern footpoint, while with $\delta_{North}=5$ for the northern
footpoint. %\textit{Many reasons to produce such difference were
%discussed until now \citep[see, for example, recent studies
%of][]{saint, zharkova2011}}.

 {Given the above note on the possible effect of the
pitch-angle anisotropy, the found mismatch between
$\delta_{South}$ and $\delta_{North}$ does not necessarily imply a
real difference in the fast electron distributions over energy in
the different footpoints, but can simply indicate inadequacy of
the adopted assumption of the distribution isotropy, which is not
too important for our study. However, this does imply } that the
fast electron energy spectrum can be harder than those derived
from the footpoints, $\delta\lesssim 4.3$. Here, unlike other
cases, we account the GS harmonic structure to give an idea of its
possible role in the total spectrum formation. The thermal plasma
contribution is negligible, since the fast electron number density
is relatively large in this event. The input (underlined italic),
and best-fit parameters are presented in Table \ref{Table 1}.

Two other events display microwave spectra overall similar to that
of the 06 January 2004 event, having thermal low-frequency parts
and nonthermal high-frequency parts. In addition, we studied the
spectrum evolution during the early stage of the 23 July 2002
event, and found that the thermal emission of an optically thick
uniform source well fits low frequency part of the observed
spectra until 00:23:45 UT. Note that the assumption of the source
homogeneity becomes incorrect at the end of the early phase
because of essential emission contribution from footpoints seen in
17 and 34 GHz images \citep{asai2009}. Perhaps it calls for three
dimensional simulations of the flaring loops
\citep{kuznetsov2011}. %Note that the interval after appearing 17
%GHz emission (from 00:23:00 to 00:23:40) was studied by
%\citet{asai2009}, who concluded that power-law indices and
%densities of the nonthermal electrons estimated from the RHESSI
%and the NoRH data differ considerably. Perhaps, assumption of the
%source homogeneity becomes incorrect at the end of the early
%phase, which calls for three dimensional simulations of the
%flaring loops \citep{kuznetsov2011}.

The spectral modeling of the  23 July 2002 yields the source
magnetic field around 200~G. Interestingly, the flare magnetic
field was independently estimated from polarization of the
optically thin free-free microwave emission \citep{asai2009} just
before the flare. The degree of the circular polarization measured
 with NoRH at 17 GHz was about 1.6\%. The degree of polarization
 is linearly proportional to the line of sight magnetic field
component. Thus, being combined with the determined magnetic field
value of 200~G, it yields an estimate of the viewing angle at the
source to be $\theta = 75^\circ$, which is reasonable for a
coronal source.  {From the RHESSI observations,
\citet{caspi2010a} and \citet{caspi2010b} have concluded that a
hot thermal plasma with temperature about 30 MK, and density
$n_{th} \geq 10^{10}$~cm$^{-3}$, was present at the beginning of
pre-impulsive phase in the corona. %This hot thermal plasma was
%cospatial with a non-thermal coronal HXR source.
The
parameters obtained from the microwave modeling for these events are given in Table 1. The estimates
from X-ray observations are available after 00:22 UT (they are given in
brackets in Table 1). The density and temperature estimates obtained from X-ray data agree well with those
of the microwave spectrum modeling. The source area increase was detected in both X-ray and microwave domains,
although the microwave source sizes were noticeably larger than the SXR source sizes. This is expected because the
SXR emission is \textit{optically thin} and so the SXR source size represents, in fact, a FWHM of the emission measure distribution, while
the microwave source area relates to \textit{optically thick} emission and so represents a \textit{full} area of the source
defined as a locus of all points corresponding to lines of sight with $\tau\ge1$. }

 {Finally, the spectrum shape of the 15 February 2011 event
is similar to the 06 January 2004 and the 23 July 2002 events.
Although, it has a noticeably higher spectrum peak frequency
indicative of correspondingly larger magnetic field, $\sim 500$~G.
For this event we used two input sets obtained, respectively, from
the GOES and RHESSI (in brackets) data.} Parameters resulted from the microwave spectral modeling are summarized in
Table \ref{Table 1}, including energy partition between the magnetic, thermal, and nonthermal energies.

\section{Discussion}

So far, the simultaneous primary heating and acceleration are
rarely studied together. It has been implicitly assumed that one
of these processes is entirely dominant over the other.
For example, it is
commonly believed that the particle acceleration is a dominating
process in a flash flare stage.
Let us discuss now what  microwave data says about the early flare phase  in addition to the commonly used X-ray
data.

First of all, in   all events under study there are nonthermal
electrons that generate gyrosynchrotron emission at frequencies
above the spectral peaks. In the case of power-law distribution of
emitting electrons the best fit indices are in the range from 2.5
up to 4, with the high-energy cutoff above 1~MeV. The densities
and energy contents of \textit{nonthermal} components (i.e., above $E_{cr}\sim10-20$~kev, see Table~1) were well below the
\textit{thermal} density and energy of the coronal sources. In two of
the studied cases the emission of nonthermal electrons
also dominates the observed spectra at frequencies below the spectrum peaks.

Secondly, we found that thermal GS emission dominates the
\textit{low-frequency} microwave spectra in many cases. This
offers reliable diagnostics of the source area and magnetic
field. The radio estimate of the {coronal} source area is
highly important because it is unbiased by the plasma density
distribution, which is unlike the SXR-derived source area. Based on the
radio data we confidently conclude that the source area grows at
the course of the flare This fully accounts for the observed
increase of the SXR-derived emission measure, while no density
increase is needed. This means that no essential chromospheric
evaporation occurs in the analyzed cases, so no energy deposition
to the chromosphere in the form of either precipitating electrons
or heat conduction takes place.

Thirdly, even though the thermal plasma contribution to the
microwave spectrum is often essential, no purely thermal stage has
been detected. Indeed,  radio signatures of the nonthermal particles appear
as soon as the plasma heating. Thus, the RHESSI (or Fermi) non-detection of
the nonthermal emission at the early flare phase is accounted by
its relatively low sensitivity, while the microwave observations
turn out to be more sensitive to small numbers of the nonthermal
electrons. Note, that in some cases (e.g., when the source area is
somewhat small, see the 13 November 2003 event as an example), the
entire microwave spectrum is dominated by the nonthermal GS
contribution.

% derived
%parameters, \textbf{the matching energy $E_{cr}$, the density of
%the nonthermal electrons with energy above $E_{cr}$, and the
%energy density ratios of various plasma components were
%estimated.}

%In particular, t
The plasma beta $\beta$ %\textbf{(the ratio of the plasma pressure to the magnetic pressure)}
is smaller than one, while the nonthermal energy density is much
smaller than the thermal one in all the cases  (i.e., $W_{nth}$
$\ll$ $W_{th}\ll W_{B}$). Although the total energy content of the
accelerated electrons is small, the available nonthermal electrons
are high efficiently accelerated from slightly nonthermal to
relativistic energies. Their spectra are hard, $\delta=2.5-3$ in
most cases, and extended up to a few MeV.  {Stated another
way, the \textit{shape} of accelerated particle spectrum at the
preflare phase is similar to that during flares, even though their
\textit{levels}
(normalizations) are highly different. %This can be understood, e.g., if the same, presumably, stochastic, acceleration mechanism,
%capable of accelerating the charged particles from somehow created \textit{'seed' population},
%is involved, though the seed populations are formed differently in  preflares or flares.
This can happen, for example, if the same, presumably stochastic,
acceleration mechanism capable of accelerating the charged
particles from somehow created \textit{``seed population''} is
involved at both preflare and flare phases. However, these seed
populations must be formed differently in preflares or flares.}

The observed significant
plasma heating suggests that the corresponding flare energy is
already available. However, it is divided highly unevenly between
the plasma heating and nonthermal seed population creation. It works in a way
similar to that in the presence of a DC electric field: at a preflare phase, a
relatively large, but still essentially sub-Dreicer field, will heat
the ambient plasma via the quasi-Joule dissipation (an enhanced, anomalous resistivity is needed to yield a significant plasma heating), while the fraction
of the runaway electrons capable of forming the mentioned seed
population, will remain relatively minor. For a larger DC electric field the
fraction of the runaway electrons will grow quickly, resulting in a
powerful nonthermal component needed to produce the impulsive flare phase.

 {Let us estimate what DC field $E$ is required to form the seed
populations at the preflare phase. Adopting typical parameters,
$n_{th}\sim 10^{10}$~cm$^{-3}$ and $T\sim30$~MK (see
Table~\ref{Table 1}), the electron Dreicer field is about
$E_{De}\approx3\times10^{-5}$~V/cm. The nonthermal to thermal
electron number density ratio is about $10^{-4}$ (see
Table~\ref{Table 1}). To build this
nonthermal component from  the maxwellian tail, electrons  with
$v>v_{cr}$ (where $\exp(-v_{cr}^2/2v_{th}^2)\sim10^{-4}$) must
runaway due to the DC electric field. Given that $v_{cr}^2\sim
(E_{De}/E)v_{th}^2$, we find $E\sim 1.5\times10^{-6}$~V/cm. Over a
typical source size of $\sim10^9$~cm, an electron can gain about
1~keV of energy. This energy is far too small compared with the
observed electron energies of  1~MeV or above. So this assumed DC
field plays no role in forming the nonthermal power-law
distribution responsible for nonthermal GS radiation. However,
this $\sim1$~keV of energy gain can  be sufficient enough to form
a slightly suprathermal seed population, from which the bulk
(presumably stochastic) acceleration produces the observed
nonthermal power-law tails up to relativistic energies.}

Microwave observations, therefore, show that the energy release
mechanism in the preflare phase is accompanied by particle
acceleration. Although this acceleration is much milder than that
in the impulsive phase, and therefore, is hard to  detect in
X-rays. The nonthermal emission produced by accelerated electrons
with energy of several hundred keV to a few MeV appears as early
as the soft X-ray emission. Thus the non-detection of the hard
X-rays at that early stage of the flares seems to originate from a
limited  sensitivity of available X-ray instruments. The frequency of the spectrum peak is
below 10 GHz for the early flare phase of microwave emission in
all cases, because of a relatively small number of accelerated
electrons at the radio sources. Microwave spectra show that
magnetic field in the coronal sources are a few hundred gauss at
the early phase. In some cases the number of accelerated electrons
is so low that the gyrosynchrotron emission from \textit{thermal} electrons
dominates the low frequency part of the microwave spectrum. In
these cases microwave
spectrum shape provides the magnetic field estimate in the coronal sources. % and cutoff energy estimation in the coronal sources. The
The microwave observations of
such events are promising for studying the transitions from the gradual
energy release to the flash flare explosives.

\section{Conclusion}

The described findings give rise to a number of fundamentally
important conclusions about the flare heating and acceleration.
(i) The flare energy release is capable of directly heating the
thermal plasma, without noticeable \textit{in situ} heating by
fast electron beams or chromospheric evaporation driven by either
electron beams or heat conduction. Since no energy transfer
process is detected, we propose the early flare phase sources to
represent the energy release and acceleration sites. (ii) The
efficient acceleration process of only a minor fraction of the
plasma electrons implies that the acceleration mechanism involved
is not capable of accelerating electrons directly from the thermal
pool, but requires a pre-extracted (injected) seed electron
population. (iii) This implies that the electron injection from
the thermal pool and their further acceleration toward higher
energies are driven by physically distinct processes. The first of
them is inefficient or somehow suppressed during the early flare
phase, while the second is already fully operational. In
particular, the acceleration by cascading turbulence alone seems
to be insufficient here. Since in the corresponding acceleration
model both injection and acceleration are driven by the same
turbulence intensity, so having inefficient injection while
efficient acceleration looks at odds to this acceleration model.
(iv) The observed significant plasma heating suggests that the
corresponding flare energy is already available. However, it is
divided highly unevenly between the plasma heating and nonthermal
population creation.  {We propose, this is due to yet
unspecified energy partition process operating} in a way
 {showing some resemblance to that controlling the balance
between Joule heating, and runaway electrons in a DC electric
field.}
%similar to that in the presence of a DC electric field: a
%relatively large but still essentially sub-Dreicer field will heat
%the ambient plasma via the Joule dissipation, while the fraction
%of the runaway electrons capable of forming the mentioned seed
%population, will remain relatively minor; for larger DC field the
%fraction of the runaway electrons will grow quickly resulting in a
%powerful nonthermal component.
We emphasize, that  {in addition to this energy partitioning
process,}   some sort of stochastic acceleration capable of
producing the observed power-law electron spectra is still needed
during both early and impulsive flare phases.

\begin{acknowledgements}
The work is supported by the Ministry of education and science of
the Russian Federation (State Contracts 16.518.11.7065 and
02.740.11.0576), and by the grants RFBR (12-02-91161-GFEN-a,
12-02-00616 and 12-02-00173-a), and by NSF grants AGS-0961867,
AST-0908344, and NASA grants NNX10AF27G and NNX11AB49G to New
Jersey Institute of Technology, and by a Marie Curie International
Research Staff Exchange Scheme Fellowship within the 7th European
Community Framework Programme. We thank Kristine Boone
from Uni. of Calgary for editing the manuscript.
\end{acknowledgements}

\bibliographystyle{apj}
\bibliography{aa_lit} %,fleishman}

%\appendix

%\section{Appendix material}

%Table
\begin{table}
\tabletypesize{\scriptsize}
%\rotate 
\caption{Parameters of radio
sources in five events. } \label{Table 1} \centering
\begin{tabular}{c c c c c c c c}
  \tableline\tableline
Parameters &  13-Nov-03 & 06-Jan-04 &
\multicolumn{3}{c}{23-Jul-02} & 20-Jan-05 & 15-Feb-11 \\
      \tableline\tableline
Time&04:57:12&06:16&00:21:00&00:22:00&00:23:45&06:30&01:47:55\\
\tableline A, $10^{18}$
cm$^{2}$&\underline{\textit{0.63}}&\underline{\textit{2.6}}&1.6&4.0\underline{\textit{(0.64)}}&16.0\underline{\textit{(0.95)}}&5.6\underline{\textit{(3.9)}}&\underline{\textit{2.1}}\\
\tableline
T, MK&\underline{\textit{25}}&\underline{\textit{35}}&\underline{\textit{32}}&\underline{\textit{32}}&\underline{\textit{32}}&18\underline{\textit{(13)}}&\underline{\textit{14(34)}}\\
\tableline
$n_{th},10^{10}$cm$^{-3}$&\underline{\textit{0.6}}&\underline{\textit{0.2}}&1.5&1.0\underline{\textit{(1.7)}}&0.2\underline{\textit{(1.7)}}&0.5\underline{\textit{(1)}}&\underline{\textit{2.0(0.3)}}\\
\tableline
$E_{cr}$, keV&15&27&27.6&29.2&26.9&4.0&12.1(15.1)\\
\tableline
B, G&70&120&200&200&200&250&500\\
\tableline
$n_{nth}$, $10^{6}$cm$^{-3}$&6.5&0.33&1.7&0.9&0.3&22&1.2(0.6)\\
\tableline
$\delta$ &2.5&2.9&3.0&2.5&2.5&4&3.0\\
\tableline
$W_{nth}/W_{th},\%$&2&0.3&0.2&0.3&0.62&1.5&0.1(0.3)\\
\tableline
$\beta$&0.2&0.03&0.08&0.06&0.01&0.01&0.008(0.003)\\
\tableline
\end{tabular}
\end{table}

\end{document}